%% file: main.tex
\documentclass[sigconf,natbib=true]{acmart}

\usepackage{multirow}

\setcopyright{none}
\acmConference[CIKM '26]
{The 35th ACM International Conference on Information and Knowledge Management}
{November 7--11, 2026}
{Rome, Italy}
\acmYear{2026}

\acmISBN{}
\acmDOI{}

\AtBeginDocument{%
  }

\begin{document}

\title{Post-Calibration Reliability Reranking of Relevance Decisions via Label-wise Monotone Projection}

\author{Inwoo Tae}
\email{vitainu0104@unist.ac.kr}
\affiliation{%
  \institution{UNIST}
  \city{Ulsan}
  \country{Republic of Korea}
}

\author{Yongjae Lee}
\email{yongjaelee@unist.ac.kr}
\affiliation{%
  \institution{UNIST}
  \city{Ulsan}
  \country{Republic of Korea}
}
\affiliation{%
  \institution{LinqAlpha}
  \city{New York}
  \state{NY}
  \country{USA}
}

\begin{abstract}
Web search, product search, and question-answering retrieval systems often assign a relevance label and confidence score to each query-candidate pair. The relevance label describes how well a page, product, or passage matches the query, while the confidence often guides downstream use or fallback decisions. Post-hoc calibration is therefore needed because misaligned confidence can make systems over-trust wrong predictions or unnecessarily defer correct ones. However, calibration mainly aligns confidence with average correctness, and does not remove predicted-label-dependent reliability differences that remain within the same calibrated confidence level. We address this gap with Label-wise Monotone Reliability Projection (MRP), which learns label-wise monotone functions that map calibrated confidence to correctness reliability while preserving the original predicted labels and class probabilities. The resulting reliability score reranks fixed predictions according to residual risk. Across six information access relevance datasets and multiple post-hoc calibrators, MRP improves reliability reranking and average fallback utility while preserving full-coverage accuracy and ECE. Structural ablations show that the main gains come from label-wise residual reliability rather than from global confidence remapping. We further analyze when MRP reliability scores can be embedded back into top-label probability geometry, showing that this projection is useful as a compatibility analysis but is distinct from the main reliability-reranking objective. The implementation will be made publicly available.
\end{abstract}


\maketitle
\vspace{-0.6em}

\section{Introduction}

Information access systems such as web search, product search, document reranking, and question answering (QA) retrieval have become central interfaces through which users discover and use information in digital services. Users search for web pages through search engines, compare products in online marketplaces, and retrieve documents or passages to support question answering. Although these services differ in form, they share a common need to evaluate multiple candidate results for a user input. In this setting, candidate evaluation can be viewed as a relevance prediction problem, where relevance refers to how well a candidate matches the user's input. These predictions then shape how candidate results are handled by the downstream system~\citep{liu2009learning,reddy2022shopping,nogueira2019passage}.

When such predictions are used downstream, the confidence attached to each decision becomes important. It may determine whether the prediction is used directly or routed to review or fallback~\citep{el2010foundations,geifman2017selective}. We use reliability for a different question from relevance: relevance is the predicted query-candidate match, whereas reliability is whether that fixed relevance decision is likely to be correct. A natural response is to apply post-hoc calibration so that confidence values better match empirical correctness frequencies~\citep{zadrozny2002transforming,guo2017calibration}. However, calibration is usually applied as an average condition over predictions with similar confidence, without directly distinguishing which relevance decision was made. It can therefore leave heterogeneous reliability within the same confidence level. As a result, a calibrated confidence group can contain different relevance decisions whose empirical correctness rates differ~\citep{gupta2021top}. Even after calibration, the system must still rerank fixed relevance decisions by residual risk among predictions with similar confidence. We view this as a post-calibration residual reliability problem, where the goal is not to produce another relevance score, but to estimate the residual correctness structure that remains after calibration.

\begin{figure*}[t]
    \centering
    \includegraphics[width=0.96\textwidth]{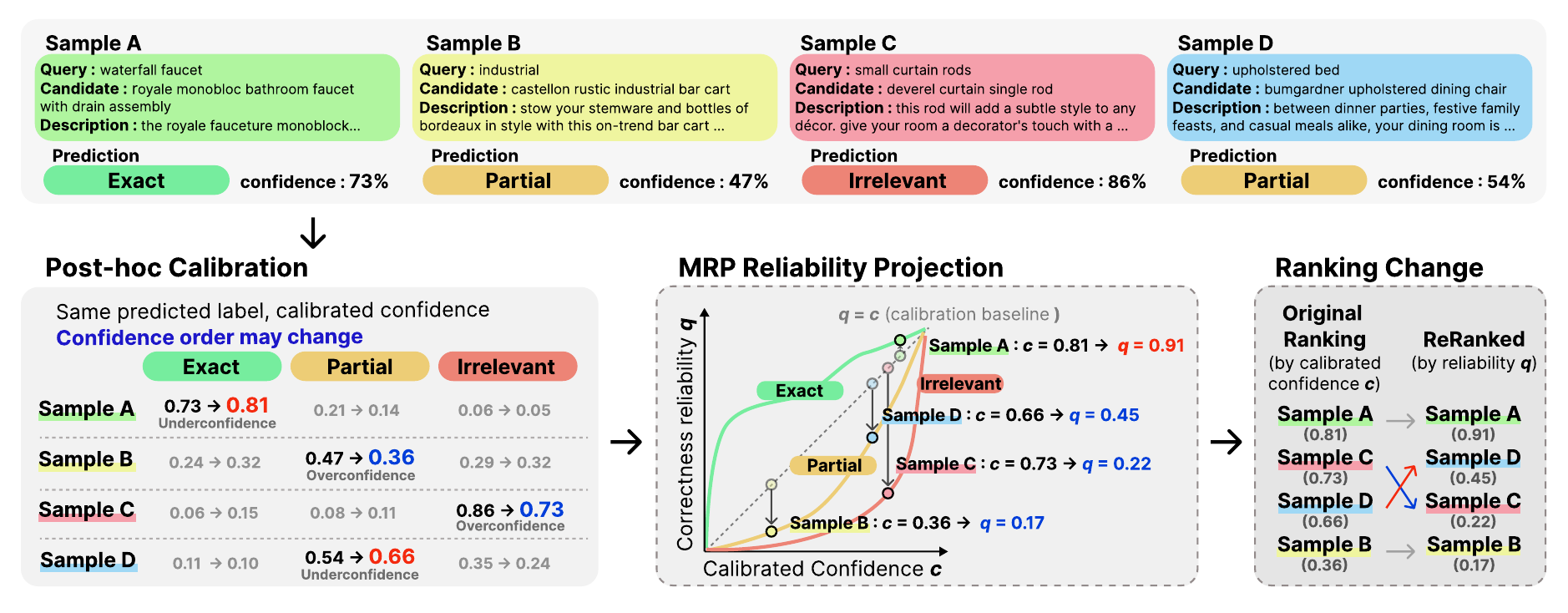}
    \caption{Overview of MRP for post-calibration reliability reranking. Each row is a fixed query-candidate relevance prediction. Post-hoc calibration updates the confidence of the predicted label, while MRP maps this calibrated confidence to label-wise correctness reliability and reranks the same predictions by estimated risk. The bottom comparison shows that confidence-based reranking and MRP reliability reranking can differ. This is reliability reranking of predictions, not retrieval ranking of candidates.}
    \Description{A method overview diagram showing query-candidate relevance decisions, post-hoc calibrated confidences, MRP reliability projection curves, and a comparison between reranking by calibrated confidence and reranking by MRP reliability. The diagram emphasizes that predicted labels remain fixed while decisions are reranked by correctness reliability rather than by candidate relevance.}
    \label{fig:mrp_method_overview}
\end{figure*}

This setting suggests a method that preserves the calibrated prediction while providing a separate score for reranking fixed decisions by expected correctness. We propose Label-wise Monotone Reliability Projection (MRP), a post-calibration reliability reranking method. Given a calibrated confidence \(c\) and a fixed predicted relevance decision \(k\), MRP learns a label-wise monotone function \(T_k(c)\) that estimates the probability that the decision is correct. Monotonicity ensures that, within the same predicted relevance decision, higher calibrated confidence does not imply lower correctness reliability. The predicted label and calibrated class probabilities remain unchanged. Only the reliability score used for reranking, review, or fallback is replaced by \(T_k(c)\). This turns calibrated confidence into a decision-conditioned correctness reliability score without turning MRP into another relevance scorer or class-probability calibrator.

We evaluate MRP on relevance prediction benchmarks covering product search, web search, question answering retrieval, e-commerce reranking, and scientific retrieval. Each experiment fixes a base relevance predictor and a post-hoc calibrator before applying MRP. The evaluation tests whether MRP improves reliability reranking while leaving class probabilities, predicted labels, full-coverage accuracy, and expected calibration error (ECE) unchanged. Across datasets and calibrators, MRP broadly improves correctness negative log-likelihood (NLL) and area under the risk-coverage curve (AURC), with AUPR-Error gains concentrated in settings where label-conditioned residual signal is present. Budgeted fallback experiments further show positive average gains in the accuracy of retained automatic predictions. This suggests that probability-scale calibration and reliability reranking capture distinct reliability aspects. Structural ablations show that this reranking gain mainly comes from residual reliability structure associated with predicted relevance decisions, rather than from confidence-only ranking. The contributions of this paper are as follows:

\begin{itemize}
    \item We define post-calibration residual reliability as a reliability reranking problem over fixed relevance decisions. This separates reliability reranking of predicted decisions from both class-probability recalibration and candidate ranking by relevance.

    \item We propose MRP, which projects calibrated confidence to correctness reliability through label-wise monotone functions. MRP preserves the calibrated class probabilities and predicted labels, and changes only the score used to rerank predictions for review, fallback, or selective use.

    \item We evaluate MRP across relevance benchmarks and post-hoc calibrators. The results show broad improvements in correctness NLL and AURC, positive average fallback gains, and AUPR-Error gains when label-conditioned residual signal is present, while preserving full-coverage accuracy and ECE.

   \item We provide structural ablations that clarify why MRP works. The main gain comes from label-conditioned residual reliability, while label-wise monotone curves provide a confidence-dependent generalization of a label-only reliability baseline.
\end{itemize}

\section{Related Work}

\subsection{Probability Calibration}

Confidence calibration began as the problem of making model scores interpretable as probabilities consistent with empirical correctness frequencies. Early post-hoc methods kept the trained predictor fixed and learned a score-to-probability mapping on a separate validation set. For example, Platt scaling uses a parametric sigmoid mapping, while histogram binning, Bayesian binning, and isotonic regression provide more flexible non-parametric mappings from prediction scores to calibrated probabilities~\citep{platt1999probabilistic,zadrozny2002transforming,niculescu2005predicting,naeini2015obtaining}.

The problem became especially prominent for deep neural networks after it was observed that accurate models can still produce overconfident probabilities. In this setting, temperature scaling became a simple yet strong post-hoc baseline that adjusts logits using a single scalar temperature~\citep{guo2017calibration}. Subsequent work has moved beyond scalar temperature by allowing more flexible transformations of class-wise logits or the full probability vector. Other lines of work directly optimize or evaluate calibration error, or distinguish full-vector calibration from top-label confidence calibration for decisions~\citep{kull2019beyond,kumar2018trainable,kumar2019verified,vaicenavicius2019evaluating,gupta2021top}.

The common goal of these methods is to make the model's class probabilities themselves more reliable as probabilities. Accordingly, they are often evaluated using calibration metrics such as expected calibration error (ECE). Our work does not replace these calibration methods. Instead, we use them as base calibrators. In our experiments, we apply the same MRP layer on top of several base calibrators, including temperature scaling, diagonal order-preserving calibration, spline-based top-label confidence calibration, h-calibration, and SMART-style calibration. Our central question is therefore not only how well a calibrator aligns the probability scale, but whether, after calibration, predictions with similar confidence can be reranked more reliably by their correctness risk.

Although MRP uses label-wise monotone functions, it is not a label-wise class-probability calibrator. A label-wise calibrator modifies the probability assigned to a class, whereas MRP keeps the calibrated probability vector fixed and learns a separate correctness-reliability score used only for reranking fixed predictions.

\subsection{Selective Prediction and Reliability Reranking}

If calibration concerns the numerical meaning of predicted probabilities, selective prediction concerns how such probabilities or uncertainty scores decide which predictions should be accepted. This line of work studies settings in which a system abstains from, rejects, or defers predictions in order to trade coverage for lower risk~\citep{el2010foundations,cortes2016learning,geifman2017selective,geifman2019selectivenet}. Related learning-to-defer work further studies how predictions can be routed to a human or expert decision maker~\citep{madras2018predict,mozannar2020consistent}. Error-detection work similarly uses scores such as maximum softmax probability to identify misclassified or out-of-distribution examples~\citep{hendrycks2016baseline}. These studies make confidence and uncertainty operational because they are not only reported as probabilities, but used as selection scores for accepting or withholding predictions.

This perspective is closely related to downstream use of calibrated probabilities, but it leaves a distinct post-calibration question. Selective prediction typically asks how much error is reduced after removing low-confidence or high-uncertainty examples. In contrast, after top-label confidence has already been calibrated, predictions with the same nominal confidence may still have different residual correctness risk. Reliability reranking focuses on this remaining problem. Before calibrated predictions are used for selection, fallback, or review, their residual risk must be estimated and used to rerank predictions without necessarily changing the calibrated class probabilities themselves.

\subsection{Relevance Prediction in Information Access Systems}

Relevance prediction is a central component of information access systems. Classical probabilistic retrieval and learning-to-rank methods use query--document relevance signals to optimize the ranking of retrieved items~\citep{robertson2009probabilistic,liu2009learning}, while neural rerankers and dense retrievers learn semantic matching signals between queries and candidate documents or passages~\citep{nogueira2019passage,karpukhin2020dense,craswell2021ms}. Similar relevance prediction problems arise in product search, where query--product pairs are annotated with graded relevance labels~\citep{reddy2022shopping,chen2022wands}, and in retrieval-augmented question answering, where retrieved passages affect downstream answer generation~\citep{guu2020retrieval,lewis2020retrieval,izacard2021leveraging}. Across these settings, relevance scores are typically evaluated by their ability to improve candidate ranking, retrieval, or classification quality.

In deployed information access systems, however, relevance scores are also used as decision signals. A system may need to decide whether to show a result directly, send a candidate to a stronger reranker, use a passage as evidence for answer generation, or defer an uncertain case for additional processing. These decisions depend not only on relative candidate ranking quality, but also on whether a relevance confidence score is reliable as a numerical signal. Existing relevance prediction work provides the scoring and candidate-ranking machinery for these systems, while calibration and selective prediction provide tools for probability interpretation and acceptance decisions. The remaining gap is the post-calibration setting in which a relevance confidence has already been assigned, but predictions with the same nominal confidence may still differ in residual correctness risk.

\section{Method}
\label{sec:method}

This section formalizes MRP as a post-calibration reliability reranking layer. We first define a fixed-decision protocol in which the base relevance decision is kept unchanged and a post-hoc calibrator assigns calibrated confidence to that decision. We then define the correctness reliability target and show why calibrated confidence alone can be insufficient when reliability depends on the predicted relevance label. Finally, we introduce the label-wise monotone reliability map, its lattice implementation, and the ablation variants used to isolate label conditioning and decision separation.

\subsection{Setup and Top-label Calibration}

Let \(x\) be an input example and \(Y\in\{1,\ldots,K\}\) its class label. Superscript \(0\) denotes the uncalibrated base predictor, and superscript \(\mathcal{A}\) denotes outputs after applying calibrator \(\mathcal{A}\). The base relevance predictor outputs \(\hat{\mathbf{p}}^{0}(x)\in[0,1]^K\), with \(\sum_{k=1}^{K}\hat{p}^{0}_{k}(x)=1\).

The top-label decision of the base predictor is fixed as
\[
    d(x)=\arg\max_k \hat{p}^{0}_{k}(x),
    \qquad
    Z(x)=\mathbf{1}\{Y=d(x)\}.
\]
Here, \(Z(x)\) indicates whether the fixed decision is correct.

A post-hoc calibrator \(\mathcal{A}\) produces \(\hat{\mathbf{p}}^{\mathcal{A}}(x)\in[0,1]^K\), with \(\sum_{k=1}^{K}\hat{p}^{\mathcal{A}}_{k}(x)=1\). Our protocol keeps \(d(x)\) fixed rather than re-evaluating the calibrated top label, so calibration cannot change decision accuracy. This isolates reliability reranking from changes in the underlying prediction. The confidence assigned by \(\mathcal{A}\) to this decision is
\[
    \hat{c}^{\mathcal{A}}(x)=[\hat{\mathbf{p}}^{\mathcal{A}}(x)]_{d(x)}
\]
where \([\mathbf{v}]_k\) denotes the \(k\)-th component of \(\mathbf{v}\). Here \(\mathcal{A}\) can be any probability-producing calibrator, including the identity map. The evaluated event is always \(Y=d(x)\).

Let \(\hat{Y}\) be a predicted top label and let \(\hat{C}\) be the confidence assigned to that prediction. The standard top-label calibration condition can be written as
\[
    \mathbb{P}(Y=\hat{Y}\mid \hat{C}=c)=c.
\]
Under our fixed-decision protocol, with \(\hat{C}^{\mathcal{A}}=\hat{c}^{\mathcal{A}}(X)\), this becomes
\[
    \mathbb{P}(Z=1\mid \hat{C}^{\mathcal{A}}=c)=c.
\]
Expected calibration error (ECE) approximates this condition over confidence bins or regions. It evaluates whether the confidence scale is correct on average, but does not imply that all predictions with the same confidence have the same correctness reliability.

\subsection{Correctness Reliability After Calibration}

The projection target is not a new class probability vector, but the probability that the selected top-label event is correct. The simplest baseline uses the calibrated score itself as the correctness reliability estimate:
\[
    \hat{q}_{\varnothing}^{\mathcal{A}}(x)=\hat{c}^{\mathcal{A}}(x).
\]
We use \(\varnothing\) to denote the case without a reliability projection. Let \(D=d(X)\) denote the random variable corresponding to the decision label selected by the uncalibrated predictor. This baseline is sufficient only if conditioning on \(D\) does not change the correctness probability beyond calibrated confidence:
\[
    H_0:\quad
    \mathbb{P}(Z=1\mid \hat{C}^{\mathcal{A}}=c,D=k)
    =
    \mathbb{P}(Z=1\mid \hat{C}^{\mathcal{A}}=c)
    \quad \forall c,k.
\]
When this equality fails, calibrated confidence alone misses label-dependent reliability differences that matter for reliability reranking. This motivates a label-wise monotone reliability map.

\subsection{Label-wise Monotone Reliability Projection}

Label-wise monotone reliability projection (MRP) estimates correctness reliability with a separate monotone function for each predicted label. For an input whose decision is \(d(x)\), MRP applies the reliability function associated with that label:
\[
    \hat{q}_{\mathrm{MRP}}^{\mathcal{A}}(x)
    =
    T_{d(x)}(\hat{c}^{\mathcal{A}}(x)),
\]
where \(T_k:[0,1]\rightarrow[0,1]\) is the confidence-to-reliability function for label \(k\). Each function satisfies the following monotonicity constraint:
\[
    c \le c'
    \quad\Longrightarrow\quad
    T_k(c) \le T_k(c')
    \qquad \forall k.
\]
This constraint prevents MRP from assigning lower reliability to a higher-confidence prediction within the same label. Thus, MRP preserves the confidence ranking within each label while learning label-specific reliability curves.

On the projection-fit set, we learn \(T_1,\ldots,T_K\) with the following objective. Let \(\mathcal{S}_{\mathrm{fit}}^{\mathcal{A}}=\{(\hat{c}_i^{\mathcal{A}},d_i,Z_i)\}_{i=1}^{m}\) be the triples produced by calibrator \(\mathcal{A}\), where \(d_i=d(x_i)\) and \(Z_i=Z(x_i)\). Let \(\mathcal{M}\) be the class of label-wise functions satisfying \(T_k:[0,1]\rightarrow[0,1]\) and monotonicity in \(c\). We use \(\mathcal{R}_{\Delta^2}(T)\) for a second-difference regularizer specified in the lattice implementation. For \(T=\{T_k\}_{k=1}^{K}\in\mathcal{M}\) and nonnegative weight \(\rho\), define
\begin{equation}
\label{eq:mrp_objective}
\begin{aligned}
    \mathcal{L}_{\rho}(T;\mathcal{S}_{\mathrm{fit}}^{\mathcal{A}})
    =
    &
    \frac{1}{m}
    \sum_{i=1}^{m}
    \mathrm{BCE}\left(Z_i,T_{d_i}(\hat{c}_i^{\mathcal{A}})\right)
    +
    \rho\,\mathcal{R}_{\Delta^2}(T).
\end{aligned}
\end{equation}
Here, \(\mathrm{BCE}(z,q)=-z\log q-(1-z)\log(1-q)\) is binary cross entropy. The BCE term fits correctness, while \(\mathcal{R}_{\Delta^2}(T)\) stabilizes finite-sample label-wise curves. MRP therefore learns a separate reliability reranking score rather than replacing the class-probability calibrator.

Although \(\hat{q}_{\mathrm{MRP}}^{\mathcal{A}}(x)\) is probability-valued, it is not a calibrated class-probability vector. It estimates only the probability that the fixed event \(Y=d(x)\) is correct. Turning it into a multiclass predictor would require allocating \(1-\hat{q}_{\mathrm{MRP}}^{\mathcal{A}}(x)\) over the non-selected labels, so we use it only as a correctness reliability score.

This distinction matters because \(\hat{p}_{d(x)}^{\mathcal{A}}(x)\) remains the base calibrator's class-probability estimate, while \(\hat{q}_{\mathrm{MRP}}^{\mathcal{A}}(x)\) is a decision-level reliability score for reranking, fallback, and review.

For fallback, selective use, and reliability reranking, predictions are ranked by decreasing estimated error probability,
\[
    1-\hat{q}_{\mathrm{MRP}}^{\mathcal{A}}(x)
    =
    1-T_{d(x)}(\hat{c}^{\mathcal{A}}(x)).
\]
Larger values indicate predictions that should be handled later, routed to fallback, or reviewed first. Importantly, MRP does not change \(\hat{\mathbf{p}}^{\mathcal{A}}(x)\), \(d(x)\), \(\hat{c}^{\mathcal{A}}(x)\), or \(Z(x)\). Therefore, full-coverage accuracy and the calibration metrics of the base calibrator remain unchanged before and after applying MRP. What changes is only how calibrated predictions are ranked for use.

\subsection{Monotone Lattice Implementation}

We do not optimize over arbitrary continuous functions \(T_k\). Instead, we represent each \(T_k\) as a finite one-dimensional monotone lattice on the confidence axis and optimize its lattice parameters. Let \(0=u_1<\cdots<u_J=1\) be knots on the confidence axis, and let \(a_{k,1},\ldots,a_{k,J}\) be the logit-scale lattice values for label \(k\). We enforce monotonicity using a nonnegative increment parameterization:
\[
    a_{k,j}
    =
    b_k+\sum_{r<j}\operatorname{softplus}(\eta_{k,r}).
\]
Here, \(\operatorname{softplus}(t)=\log(1+\exp(t))\), so this parameterization guarantees \(a_{k,1}\le\cdots\le a_{k,J}\). For \(c\in[u_j,u_{j+1}]\), we interpolate on the logit scale:
\[
    \tilde{a}_k(c)
    =
    \frac{u_{j+1}-c}{u_{j+1}-u_j}a_{k,j}
    +
    \frac{c-u_j}{u_{j+1}-u_j}a_{k,j+1}.
\]
The reliability map is then \(T_k(c)=\sigma(\tilde{a}_k(c))\), where \(\sigma(t)=1/(1+\exp(-t))\). Thus, the parameterization satisfies both \(0\le T_k(c)\le1\) and monotonicity throughout optimization. The second-difference regularizer in Eq.~\eqref{eq:mrp_objective} is
\[
    \mathcal{R}_{\Delta^2}(T)
    =
    \frac{1}{K(J-2)}
    \sum_{k=1}^{K}\sum_{j=2}^{J-1}
    \left(a_{k,j+1}-2a_{k,j}+a_{k,j-1}\right)^2,
\]
which penalizes highly oscillatory label-wise curves.

At test time, each fixed-decision pair \((\hat{c}_i^{\mathcal{A}},d_i)\) receives
\[
    \hat{q}_{\mathrm{MRP},i}^{\mathcal{A}}=T_{d_i}(\hat{c}_i^{\mathcal{A}}),
\]
and predictions are ranked by \(1-\hat{q}_{\mathrm{MRP},i}^{\mathcal{A}}\).

\subsection{Ablation Variants}
\label{sec:ablation_variants}

The main MRP model above, the label-wise 1D variant, learns one monotone curve \(T_k(c)\) for each predicted label. We compare it with variants that remove or add one structural component. Together with the confidence baseline \(\hat{q}_{\varnothing}^{\mathcal{A}}\), these variants separate three effects, namely using confidence alone, conditioning on the predicted label, and allowing the label effect to vary with confidence.

\input{table/datasets} 

Shared 1D \(T(c)\) removes label conditioning and learns a single global confidence-to-reliability curve. Throughout this subsection, \(\operatorname{logit}(p)=\log(p/(1-p))\). The label-only intercept baseline removes confidence-dependent curves but keeps a learned label intercept:
\[
    \hat{q}_{\mathrm{LI}}^{\mathcal{A}}(x)
    =
    \sigma\left(\operatorname{logit}(\hat{c}^{\mathcal{A}}(x))+\alpha_{d(x)}\right)
\]
where \(\alpha_k\) is learned for each predicted label. This diagnostic variant tests whether the observed gain can be explained by a fixed label-wise intercept. It is a special case of \(T_k(c)\) where each label applies a fixed logit-scale shift rather than a confidence-dependent reliability curve. We also include per-label isotonic projection, which fits an independent monotone confidence-to-reliability map for each predicted label without the smooth lattice parameterization used by MRP.

Label-wise 2D adds decision separation to the main label-wise model. Let the uncalibrated runner-up label be
\[
    \ell(x)=\arg\max_{k\neq d(x)}\hat{p}^{0}_{k}(x)
\]
and define the calibrated top-runner logit gap as
\[
    g^{\mathcal{A}}(x)
    =
    \operatorname{logit}([\hat{\mathbf{p}}^{\mathcal{A}}(x)]_{d(x)})
    -
    \operatorname{logit}([\hat{\mathbf{p}}^{\mathcal{A}}(x)]_{\ell(x)}).
\]
The 2D variant learns
\[
    \hat{q}_{\mathrm{2D}}^{\mathcal{A}}(x)=T_{d(x)}(\hat{c}^{\mathcal{A}}(x),g^{\mathcal{A}}(x))
\]
where \(T_k(c,g)\) is constrained to be nondecreasing in both \(c\) and \(g\). This variant tests whether top-runner separation provides additional reliability information beyond confidence and the fixed decision label.

\section{Experiments}

MRP is evaluated on relevance prediction tasks spanning product search, web search, QA retrieval, e-commerce reranking, and scientific document retrieval. All experiments follow the fixed-decision protocol from Section~\ref{sec:method}, where base decisions are held fixed, post-hoc calibrators only change the confidence assigned to those decisions, and MRP only reranks calibrated predictions by reliability. The main question is whether this improves correctness reliability reranking while preserving full-coverage predictions and the calibrated probabilities.

\subsection{Datasets}

The experiments use six datasets that cover both graded and binary relevance decisions. Amazon ESCI and WANDS represent product search, MSLR-WEB10K represents graded web relevance, Alloprof-Rerank and SciDocs represent retrieval reranking, and ESCI-Rerank-US provides a binary e-commerce reranking setting. These datasets differ in domain and label granularity, but in all cases the predicted relevance label can affect whether a candidate is trusted for downstream use. Table~\ref{tab:datasets} summarizes the datasets used in our experiments. \(K\) denotes the number of relevance classes.

\subsection{Base Predictors and Calibrators}

For each dataset, we first train a base relevance predictor and then fit post-hoc calibrators on validation data. For text-based query--candidate datasets, the predictor is a lightweight linear classifier over the concatenated query and candidate text. For feature-based learning-to-rank data, we use the provided ranking features. The goal of this setup is not to maximize the base relevance model itself, but to create a fixed prediction source on which calibration and post-calibration reliability reranking can be compared cleanly.

The main comparison starts from a single fixed base predictor. Concretely, for each dataset we use the saved logits from model seed \(0\) and define the fixed decision \(d(x)\) from this uncalibrated predictor. All calibrators are then applied to the same saved logits and only change the probability or confidence assigned to this fixed decision. This fixed-decision protocol ensures that calibration methods and MRP share the same decision event \(Y=d(x)\), so full-coverage accuracy is identical within each dataset.

The comparison includes six calibration outputs. Uncal. uses the identity map and keeps softmax probabilities. TS~\citep{guo2017calibration} fits a scalar temperature to rescale logits. DIAG~\citep{rahimi2020intra} follows the diagonal intra-order-preserving calibration family. For fixed-decision evaluation, it rescales sorted adjacent logit gaps with positive factors to preserve the within-sample class order. Spline~\citep{gupta2020calibration} applies spline-based top-label confidence calibration. h-cal~\citep{huang2025h} uses the h-calibration objective. SMART~\citep{guo2025sample} follows sample margin-aware recalibration of temperature scaling and predicts a sample-wise temperature from the top-two logit gap. For all outputs, MRP leaves \(\hat{\mathbf{p}}^{\mathcal{A}}(x)\), \(d(x)\), and \(\hat{c}^{\mathcal{A}}(x)\) unchanged and estimates only the correctness reliability of the fixed decision.

\subsection{Implementation Details}

For each protocol seed, we split validation data into calibrator-fit, projection-fit, and projection-selection subsets. The calibrator-fit subset fits the base calibrator \(\mathcal{A}\). The projection-fit subset is used to learn \(T_k\) or an ablation variant, and the projection-selection subset is held out for model selection in variants that require it. For Uncal., the base calibrator is the identity map and therefore has no fitted calibration parameters, but we still repeat the projection-fit and projection-selection split to estimate MRP under the same three-seed protocol. The test set is evaluated only after the calibrator, reliability projection, and hyperparameters are fixed.

\input{table/main_mrp_table} 

The main MRP model is implemented as the label-wise monotone 1D lattice \(T_k(\hat{c}^{\mathcal{A}})\) described in Section~\ref{sec:method}. Unless otherwise stated, we use \(J=8\) confidence knots, second-difference coefficient \(\rho=10^{-4}\), the Adam optimizer with learning rate \(0.03\), and a maximum of 400 epochs. To keep computation comparable across datasets, projection-fit and projection-selection are each capped at 8000 samples. The base predictor is trained or loaded once with model seed \(0\), and the calibrator and MRP procedures are repeated with protocol seeds \(1,2,3\). For fitted calibrators, this repeats both the calibration fit and the MRP fit. For Uncal., only the MRP fit varies because the identity calibrator is deterministic. For every run, we verify that the reported full-coverage accuracy matches the seed-0 uncalibrated decision accuracy, confirming that calibration and MRP do not change the evaluated decision event. We report means and standard deviations over the protocol seeds.

The ablation variants defined in Section~\ref{sec:method} use the same data split, hyperparameter-selection protocol, and evaluation procedure. All experiments are implemented in PyTorch 2.1.0 with CUDA 12.1. GPU runs use a single NVIDIA RTX A5000 24GB GPU. Calibration and MRP fitting are performed on saved logits, so every calibrator and MRP comparison shares the same prediction source unless the base relevance model is explicitly retrained.

\subsection{Evaluation Metrics}

We evaluate the class-probability quality of the base calibrator with full-coverage accuracy, ECE, multiclass NLL, and Brier score. Since MRP does not modify \(\hat{\mathbf{p}}^{\mathcal{A}}\), these quantities must remain identical before and after applying MRP. The main table therefore first reports that Acc. and ECE are preserved, and then evaluates MRP with metrics for correctness-reliability reranking.

MRP is evaluated through the correctness reliability estimate \(\hat{q}_i^{\mathcal{A}}\) and the corresponding estimated error probability \(1-\hat{q}_i^{\mathcal{A}}\). Correctness negative log-likelihood, denoted as \(\mathrm{NLL}_{\mathrm{correct}}\), is defined as
\[
    \mathrm{NLL}_{\mathrm{correct}}
    =
    -\frac{1}{n}\sum_{i=1}^{n}
    \left[
    Z_i\log \hat{q}_i^{\mathcal{A}}
    +
    (1-Z_i)\log(1-\hat{q}_i^{\mathcal{A}})
    \right]
\]
and measures how well \(\hat{q}_i^{\mathcal{A}}\) predicts the probability that the fixed decision is correct. Lower values are better.

AUPR-Error follows failure-prediction and misclassification-detection evaluations~\citep{hendrycks2016baseline,corbiere2019addressing}. It is the standard area under the precision--recall curve (AUPR), computed for detecting incorrect predictions rather than relevant items. Let \(W_i=1-Z_i\) be the wrong-prediction indicator, let \(s_i=1-\hat{q}_i^{\mathcal{A}}\) be the estimated error score, and let \(\pi\) rank examples by decreasing \(s_i\). We compute
\[
    \mathrm{AUPR\mbox{-}Error}
    =
    \frac{1}{\sum_i W_i}
    \sum_{j=1}^{n}
    W_{\pi_j}
    \frac{\sum_{t=1}^{j} W_{\pi_t}}{j}.
\]
It measures whether predictions flagged as high risk are actually wrong.

AUPR-Error is important for selective prediction and fallback routing, where the operational goal is often to prioritize likely failures for review or fallback rather than to change every predicted probability. Higher values are better.

Area under the risk-coverage curve, or AURC, measures the area under the coverage-risk curve obtained by accepting predictions from low risk to high risk. Given a ranking \(\pi\) by increasing estimated risk,
\[
    \mathrm{AURC}
    =
    \frac{1}{n}
    \sum_{j=1}^{n}
    \frac{1}{j}
    \sum_{t=1}^{j}
    (1-Z_{\pi_t})
\]
and lower values are better.

Finally, we report selective accuracy to simulate a budgeted fallback setting. At coverage \(\tau\), the system automatically handles the lowest-risk \(\tau n\) predictions and routes the remaining \(1-\tau\) fraction to fallback or review. We define
\[
    \mathrm{SelAcc}@\tau
    =
    \frac{1}{|\mathcal{R}_{\tau}|}
    \sum_{i\in\mathcal{R}_{\tau}} Z_i
\]
where \(\mathcal{R}_{\tau}\) is the set of retained predictions with the lowest estimated risk. This metric tests whether MRP places usable predictions earlier under a limited fallback budget. We use \(\tau\in\{0.1,0.5,0.7,0.9\}\) to examine whether the reranking gain remains stable across different fallback budgets.

\section{Results and Analysis}

\subsection{Main Reliability Results}

The first analysis asks whether MRP addresses the intended reliability task after calibration. The evaluation has two questions. First, does MRP preserve the class probabilities and full-coverage predictions produced by the base calibrator? Second, without changing these probabilities, does it rerank calibrated predictions by correctness reliability more effectively? Table~\ref{tab:main_mrp} reports both aspects.

For any base calibrator \(\mathcal{A}\), MRP leaves \(\hat{\mathbf{p}}^{\mathcal{A}}(x)\), \(d(x)\), \(\hat{c}^{\mathcal{A}}(x)\), and \(Z(x)\) unchanged. Therefore, Acc. and ECE must be identical before and after applying MRP, and the table reports these quantities once for each base calibrator. In contrast, \(\mathrm{NLL}_{\mathrm{correct}}\), AUPR-Error, and AURC compare the no-projection confidence baseline \(\hat{q}_{\varnothing}^{\mathcal{A}}\) with the MRP reliability score \(\hat{q}_{\mathrm{MRP}}^{\mathcal{A}}\).

Across the evaluated settings, MRP improves the main correctness-reliability metrics on average while leaving Acc. and ECE unchanged. The largest gains appear when calibrated confidence alone is a weak reranking signal, as in MSLR-WEB10K and Alloprof-Rerank. These improvements are not changes in prediction accuracy. They mean that the same fixed predictions are ranked more usefully for identifying likely errors.

The table also shows why this task is different from ordinary calibration. Some calibrators substantially improve ECE but leave almost the same confidence-baseline reliability reranking. ESCI-Rerank-US is a clear example. TS, DIAG, Spline, and SMART sharply reduce ECE, but their confidence-baseline AUPR-Error remains nearly unchanged. After MRP, AUPR-Error improves for the same fixed decisions. Thus, probability calibration improves the scale of confidence, whereas MRP improves how calibrated predictions are prioritized by residual correctness risk.

The gains are not uniform in every metric. SciDocs improves in \(\mathrm{NLL}_{\mathrm{correct}}\) and AURC, but AUPR-Error is nearly unchanged or slightly lower. This exception clarifies the scope of the method. MRP does not uniformly improve every risk metric. It is most effective when label-conditioned residual reliability provides a usable reranking signal. Even here, the reduction in AURC indicates that the retained low-risk prefix becomes better ranked, while error detection by AUPR remains difficult.

\begin{figure*}[htbp]
    \centering
    \includegraphics[width=0.96\textwidth]{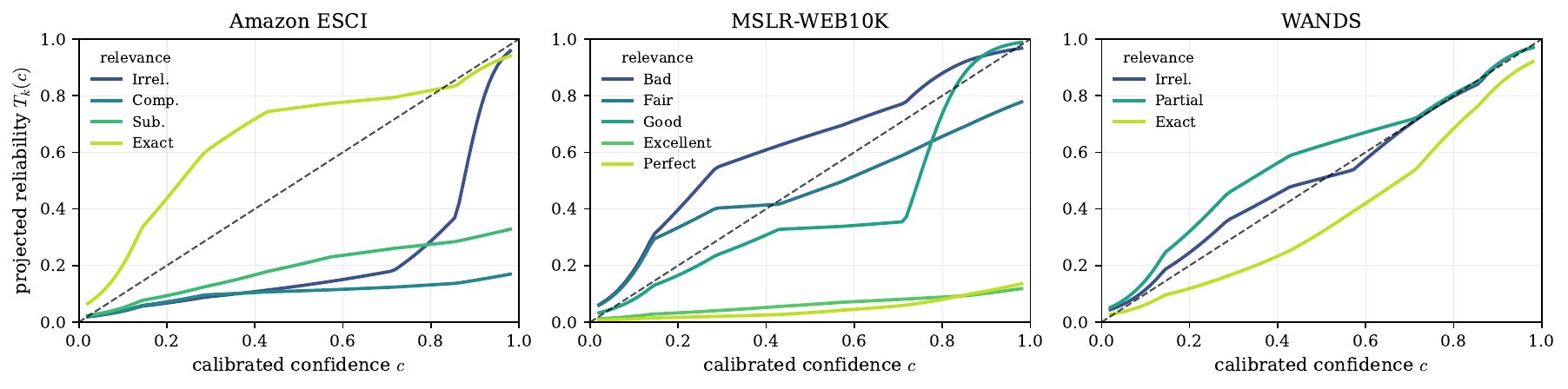}
    \caption{Learned label-wise reliability curves \(T_k(c)\), shown under SMART calibration as a representative base calibrator. The dashed line is the identity \(q=c\). The curves show that equal calibrated confidence can correspond to different correctness reliability depending on the predicted relevance label.}
    \Description{A multi-panel line plot showing label-wise monotone reliability curves for several datasets. Each panel compares learned label-specific curves against a dashed identity line, illustrating that the same calibrated confidence can be mapped to different correctness reliability values depending on the predicted label.}
    \label{fig:mrp_labelwise_curves}
\end{figure*}

\subsection{Budgeted Fallback Performance}

Table~\ref{tab:budgeted_fallback} gives a more operational view of the same reliability reranking problem. It reports selective-accuracy gains over the confidence baseline at coverage \(\tau\). Here, a system routes the highest-risk predictions to fallback or review and automatically handles only the retained lowest-risk predictions. The goal is not to change the predicted probabilities, but to decide which calibrated predictions should remain in automatic use under a limited fallback budget.

Positive values mean that MRP keeps a more accurate retained set than the confidence baseline. Average gains are positive at every coverage level and largest when few predictions can be used automatically. This is where reliability reranking matters most, because few can be accepted without fallback.

The dataset-level pattern is also informative. MSLR-WEB10K shows the strongest gains, while ESCI-Rerank-US and Alloprof-Rerank also improve at moderate coverages. Amazon ESCI and WANDS show smaller gains, and SciDocs is close to neutral after the most selective setting. This matches Table~\ref{tab:main_mrp}, where SciDocs had little AUPR-Error gain.

Overall, the fallback simulation gives MRP an operational interpretation. The method does not claim that class probabilities change or that the model becomes more accurate at full coverage. It shows that, when only a fraction can be handled automatically, the same calibrated predictions can be ranked more usefully.

\input{table/budgeted_fallback} 

\input{table/label_conditional_residual} 

\subsection{Evidence for Label-wise Residual Reliability}

We next test whether confidence alone is sufficient for reliability. If true, residual reliability would be similar across predicted labels within each confidence region.

To diagnose this effect, we group predictions by confidence and measure how much the average residual \(Z-\hat{c}^{\mathcal{A}}\) varies across predicted labels. For a confidence group \(B\), we define
\[
\begin{aligned}
    \mathrm{Spread}(B)
    &=
    \max_k
    \mathbb{E}[Z-\hat{c}^{\mathcal{A}} \mid X\in B, D=k] \\
    &\quad -
    \min_k
    \mathbb{E}[Z-\hat{c}^{\mathcal{A}} \mid X\in B, D=k].
\end{aligned}
\]
If this quantity is close to zero, then the residual reliability is nearly the same across predicted labels within the same confidence group. A large value means that two predictions with similar calibrated confidence can have different residual errors depending on which label was predicted.
Table~\ref{tab:label_conditional_residual} reports the average label spread across confidence groups for SMART as a representative calibrator. The random column shows how large the spread would be if predicted labels were shuffled inside each confidence group. In every dataset, the label spread is larger than this random baseline, and the pattern appears in all three protocol seeds. Thus, the confidence-only assumption from Section~\ref{sec:method} fails empirically. Residual reliability remains label-dependent within confidence groups.

The spread explains why label-wise reliability functions are useful. MSLR-WEB10K, Amazon ESCI, and Alloprof-Rerank show the largest spreads, far above their shuffled-label baselines. These are also the datasets where the reliability reranking gains in Table~\ref{tab:main_mrp} are large. ESCI-Rerank-US has a smaller but clear spread, while SciDocs has the weakest signal. This is consistent with the weaker AUPR-Error improvement above.

Figure~\ref{fig:mrp_labelwise_curves} visualizes the learned functions \(T_k(c)\). The dashed identity line \(q=c\) uses base confidence directly. The learned curves show that the same calibrated confidence can map to different reliability values depending on the predicted label. This illustrates that MRP is not merely smoothing confidence again. It estimates separate correctness reliability curves for different decision labels.

\input{table/mrp_ablation_results} 

\subsection{Structural Ablations}

Table~\ref{tab:mrp_ablation_results} shows which structural assumption is responsible for the gain. Shared 1D improves \(\mathrm{NLL}_{\mathrm{correct}}\) over the confidence baseline, but it does not improve AUPR-Error, AURC, or SelAcc@50. This means that simply relearning a shared confidence-to-correctness curve can improve probability fit, but it does not improve risk reranking.

These variants form a nested diagnostic sequence. Shared 1D tests for a global confidence-to-correctness mismatch. The label-only intercept tests whether the gain can be explained by static predicted-label bias, and per-label isotonic tests a nonparametric label-wise monotone mapping. MRP then asks whether the same label-conditioned structure can be represented by smooth monotone reliability curves.

The reranking metrics improve only after the decision label is introduced. The strong performance of the label-conditioned variants should not be read as evidence against MRP. Rather, it identifies the main residual structure that MRP is designed to model. Much of the residual signal after calibration is label-conditioned. MRP, the label-wise 1D model, provides a smooth monotone implementation of this idea by learning \(T_k(c)\) for each label. Compared with per-label isotonic projection, this smooth lattice gives a clearer \(\mathrm{NLL}_{\mathrm{correct}}\) improvement while maintaining comparable reranking quality. Its gains over label-only and per-label isotonic variants are modest, so the result should not be read as MRP dominating all label-wise alternatives. Instead, it shows that the key ingredient is label-conditioned residual reliability, with MRP serving as the main monotone projection used in this paper.

\input{table/simplex_projection} 

Label-wise 2D, which uses the top-runner gap, is nearly identical to Label-wise 1D. Its \(\mathrm{NLL}_{\mathrm{correct}}\) is slightly lower, but AUPR-Error, AURC, and SelAcc@50 do not improve. This supports the final design choice. The main residual structure in these relevance datasets is the label-wise confidence-to-reliability curve \(T_k(c)\), not an additional two-dimensional decision-gap surface. We therefore use label-wise 1D projection as the main method.

As a robustness check, we test whether this signal is specific to the lightweight linear base predictor. On MSLR-WEB10K, a balanced LightGBM multiclass relevance predictor raises base accuracy from 0.394 to 0.518 and reduces ECE from 0.292 to 0.014. Even with this stronger and well-calibrated base, MRP improves the confidence baseline from 0.629 to 0.586 in \(\mathrm{NLL}_{\mathrm{correct}}\), from 0.631 to 0.730 in AUPR-Error, and from 0.318 to 0.286 in AURC. This reinforces the same conclusion as Table~\ref{tab:mrp_ablation_results}: the persistent signal is label-conditioned residual reliability, and MRP is the smooth monotone projection used as our main variant.

\section{Reliability-to-Probability Projection}

\subsection{Simplex View of Top-label Projection}

A class probability vector is a point on the probability simplex. The vertex \(\mathbf e_k\) assigns probability one to class \(k\). For a fixed decision \(d(x)\), the simplex center has top-label probability \(1/K\), while the decision-label vertex \(\mathbf e_{d(x)}\) has top-label probability one. Thus, changing the confidence assigned to the fixed decision can be viewed as moving a probability vector toward or away from that vertex.

Temperature or power transforms trace an order-preserving path from the uniform distribution toward the argmax-class vertex. Related simplex views of calibrated predictive distributions are also used in recent uncertainty-calibration work~\citep{heiss2026jucal}. This matters because MRP estimates a probability-valued scalar,
\[
    q_{\mathrm{MRP}}^{\mathcal{A}}(x)=T_{d(x)}(\hat{c}^{\mathcal{A}}(x)),
\]
for the correctness event \(Y=d(x)\), but uses it for reliability reranking rather than as a multiclass probability vector.

For a probability vector \(\mathbf p\), the power-normalized path
\[
    p_k(\alpha)=\frac{p_k^\alpha}{\sum_j p_j^\alpha}
\]
gives a one-dimensional trajectory. At \(\alpha=1\), it returns the original vector. As \(\alpha\rightarrow0\), it approaches the simplex center, and as \(\alpha\rightarrow\infty\), it concentrates on the largest-probability class. When \(d(x)\) remains the top class, this path sweeps the decision-label probability from \(1/K\) toward 1 without changing the class order.

This view lets us ask whether \(q_{\mathrm{MRP}}^{\mathcal{A}}(x)\) can be realized as the top-label probability of a decision-preserving class probability vector. We define this optional extension as MRP-induced reliability calibration (MRC), a compatibility analysis between the learned reliability space and class-probability geometry.

\subsection{MRC Construction}

MRC starts from the calibrated vector \(\hat{\mathbf p}^{\mathcal{A}}(x)\) and the MRP target \(q_{\mathrm{MRP}}^{\mathcal{A}}(x)\). It moves \(\hat{\mathbf p}^{\mathcal{A}}(x)\) along a sample-wise power-temperature path so that the fixed top-label probability matches the target as closely as possible:
\[
    \hat p^{\mathcal{A},\alpha}_k(x)
    =
    \frac{(\hat p^{\mathcal{A}}_k(x))^{\alpha}}
    {\sum_j(\hat p^{\mathcal{A}}_j(x))^{\alpha}}
\]
which is equivalent to sample-wise temperature scaling of \(\log \hat{\mathbf p}^{\mathcal{A}}(x)\). If \(d(x)\) is the top label and \(q_{\mathrm{MRP}}^{\mathcal{A}}(x)\in[1/K,1]\), we can find \(\alpha_i\ge 0\) such that
\[
    \hat p^{\mathcal{A},\alpha_i}_{d(x_i)}(x_i)=q_{\mathrm{MRP}}^{\mathcal{A}}(x_i).
\]
When an exact solution exists, MRC chooses it. Otherwise, it uses the closest feasible boundary point on the same path. With selected parameter \(\alpha_i^\star\),
\[
    \hat{\mathbf p}^{\mathrm{MRC}}(x_i)=\hat{\mathbf p}^{\mathcal{A},\alpha_i^\star}(x_i).
\]

\begin{figure}[t]
    \centering
    \includegraphics[width=0.8\linewidth]{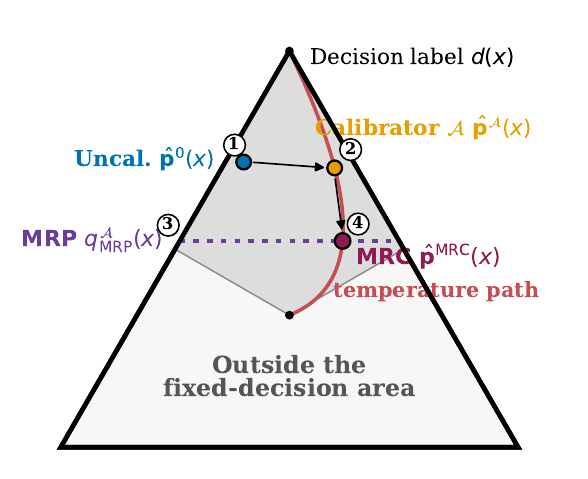}
    \caption{Schematic of MRC on the probability simplex. The numbered markers show the sequence from the uncalibrated prediction to the calibrated prediction, the MRP target contour \(p_{d(x)}=q_{\mathrm{MRP}}^{\mathcal{A}}(x)\), and the MRC projection. The projection moves along the sample-wise temperature path and realizes the MRP target as a fixed top-label probability when feasible.}
    \Description{A triangular probability-simplex diagram with numbered markers. The diagram shows an uncalibrated point, a calibrated point, a dashed MRP target contour, and an MRC projected point on a curved temperature path toward the decision-label vertex.}
    \label{fig:mrc_simplex_schematic}
\end{figure}

Unlike standard TS, MRC does not learn one global temperature. It chooses a sample-wise endpoint from \(q_{\mathrm{MRP}}^{\mathcal{A}}(x)\). If \(q_{\mathrm{MRP}}^{\mathcal{A}}(x)<1/K\), the target cannot be represented exactly on this path. If \(d(x)\) is no longer the top label after calibration, the decision-preserving path is not directly applicable. We report feasibility with the projection results. Figure~\ref{fig:mrc_simplex_schematic} illustrates the process.

\subsection{Projection Results and Feasibility}

Table~\ref{tab:simplex_projection} tests whether MRP reliability estimates fit top-label probability geometry. MRC finds a sample-wise \(\alpha_i^\star\) on the power-temperature path and compares the resulting vector with the calibrated one. This is stricter than reliability reranking because the score must be representable as a top-label probability.

The results are mixed but informative. MRC often improves NLL and Brier score, and it can reduce ECE when base probabilities are poorly calibrated. However, it does not uniformly improve fixed ECE or top-label ECE. For strong calibrators, the projection can move away from the best calibration point. Thus, probability correction is more constrained than reliability reranking.

The feasibility columns make this point concrete. In Amazon ESCI and WANDS, only 7.2\% and about 1.0\% of predictions have \(q_{\mathrm{MRP}}<1/K\), so most targets are feasible. In ESCI-Rerank-US, MSLR-WEB10K, and Alloprof-Rerank, this rate is larger, around 25--38\%, so many low-reliability predictions hit the simplex boundary. Thus, MRC is best viewed as a compatibility analysis of whether the MRP reliability space embeds into probability geometry.

The central contribution remains MRP: a label-wise correctness-reliability space for reranking predictions after calibration. MRC shows that this space can sometimes be embedded through constrained top-label projection, but it should not be read as a replacement for standard multiclass calibration.

\section{Limitations}

MRP is designed for post-calibration reliability reranking, not for improving the underlying relevance model or replacing a multiclass calibrator. Because it preserves the predicted relevance label and calibrated class probabilities, it cannot fix errors that require changing the model decision. The fixed-decision protocol is useful when systems must decide which existing predictions to trust, but it does not address candidate generation, retrieval quality, or end-to-end relevance ranking.

The method is most useful when residual reliability differs across predicted labels after calibration. When this structure is weak, as in SciDocs, confidence already captures much of the available signal and MRP has less room to improve reranking. MRC is only a secondary analysis: not all correctness-reliability estimates are feasible top-label probabilities. Extending MRP to query-dependent reliability functions, richer decision contexts, and joint optimization with candidate ranking remains future work.

\section{Conclusion}

We study residual reliability reranking after calibration in relevance prediction and introduce MRP, a label-wise monotone projection method. Even when calibrated probabilities are fixed, predictions with similar confidence can have different correctness reliability depending on the predicted relevance label. Instead of changing class probabilities, MRP learns label-wise correctness-reliability curves and reranks fixed predictions by residual risk.

Across six information access datasets and multiple post-hoc calibrators, MRP broadly improves \(\mathrm{NLL}_{\mathrm{correct}}\) and AURC, improves fallback performance on average, and improves AUPR-Error when label-conditioned residual signal is present, while preserving full-coverage accuracy and ECE. The ablations show that the main signal comes from label-wise residual reliability rather than from a shared confidence remapping or an additional decision-gap dimension.
These results suggest that calibration does not end the reliability problem. Calibrated confidence gives a meaningful probability scale, but systems that must decide which predictions to use, defer, or review still need reliability reranking.

\bibliographystyle{ACM-Reference-Format}
\bibliography{references}

\end{document}

%% file: table/datasets.tex
\begin{table*}[t]
\centering
\small
\caption{Datasets used in the relevance prediction experiments.}
\label{tab:datasets}
\begin{tabular}{llcll}
\toprule
Dataset & Domain & \(K\) & Labels & Task \\
\midrule
Amazon ESCI~\citep{reddy2022shopping}
& E-commerce search
& 4
& Exact/Substitute/Complement/Irrelevant
& Query-product relevance classification \\
\addlinespace[2pt]
MSLR-WEB10K~\citep{microsoft2010mslr}
& Web search
& 5
& Relevance grades 0--4
& Learning-to-rank relevance prediction \\
\addlinespace[2pt]
Alloprof-Rerank~\citep{lefebvre2023alloprof}
& QA retrieval
& 2
& Relevant/Non-relevant
& Question-document relevance reranking \\
\addlinespace[2pt]
ESCI-Rerank-US~\citep{reddy2022shopping}
& E-commerce reranking
& 2
& Relevant/Non-relevant
& Query-product candidate reranking \\
\addlinespace[2pt]
WANDS~\citep{chen2022wands}
& E-commerce search
& 3
& Exact/Partial/Irrelevant
& Product search relevance prediction \\
\addlinespace[2pt]
SciDocs~\citep{cohan2020specter,thakur2021beir}
& Scientific retrieval
& 2
& Relevant/Non-relevant
& Candidate document reranking \\
\bottomrule
\end{tabular}
\end{table*}

%% file: table/main_mrp_table.tex
\begin{table*}[t]
\centering
\scriptsize
\caption{Post-calibration reliability reranking results under the fixed-decision protocol. Acc. and ECE are base-calibrator values and are unchanged by MRP. Reliability metrics compare the confidence baseline \(\hat q=\hat c^{\mathcal{A}}\) with MRP. Blue bold \(\Delta\) values indicate improvements in the direction of each metric.}
\label{tab:main_mrp}
\setlength{\tabcolsep}{3.2pt}
\newcommand{\improve}[1]{\textcolor{blue}{\textbf{#1}}}
\begin{tabular}{llcc|ccc|ccc|ccc}
\toprule
\multirow{2}{*}{Dataset} & \multirow{2}{*}{Calibrator} & \multirow{2}{*}{Acc.} & \multirow{2}{*}{ECE} & \multicolumn{3}{c|}{\(\mathrm{NLL}_{\mathrm{correct}}\downarrow\)} & \multicolumn{3}{c|}{AUPR-Error$\uparrow$} & \multicolumn{3}{c}{AURC$\downarrow$} \\
 & & & & Conf. & MRP & $\Delta$ & Conf. & MRP & $\Delta$ & Conf. & MRP & $\Delta$ \\
\midrule
\multirow{6}{*}{\shortstack{ESCI-Rerank\\US}} & Uncal. & 0.571 & 0.183 & 0.832{\tiny $\pm$ 0.000} & 0.647{\tiny $\pm$ 0.000} & \improve{-0.186} & 0.462{\tiny $\pm$ 0.000} & 0.558{\tiny $\pm$ 0.000} & \improve{+0.097} & 0.385{\tiny $\pm$ 0.000} & 0.331{\tiny $\pm$ 0.000} & \improve{-0.053} \\
 & TS & 0.571 & 0.013 & 0.679{\tiny $\pm$ 0.000} & 0.646{\tiny $\pm$ 0.000} & \improve{-0.033} & 0.462{\tiny $\pm$ 0.000} & 0.558{\tiny $\pm$ 0.000} & \improve{+0.097} & 0.385{\tiny $\pm$ 0.000} & 0.331{\tiny $\pm$ 0.000} & \improve{-0.053} \\
 & DIAG & 0.571 & 0.011 & 0.679{\tiny $\pm$ 0.000} & 0.646{\tiny $\pm$ 0.000} & \improve{-0.033} & 0.462{\tiny $\pm$ 0.000} & 0.558{\tiny $\pm$ 0.000} & \improve{+0.097} & 0.385{\tiny $\pm$ 0.000} & 0.332{\tiny $\pm$ 0.000} & \improve{-0.053} \\
 & Spline & 0.571 & 0.015 & 0.680{\tiny $\pm$ 0.000} & 0.646{\tiny $\pm$ 0.000} & \improve{-0.034} & 0.466{\tiny $\pm$ 0.003} & 0.563{\tiny $\pm$ 0.002} & \improve{+0.097} & 0.387{\tiny $\pm$ 0.001} & 0.332{\tiny $\pm$ 0.001} & \improve{-0.055} \\
 & h-cal & 0.571 & 0.112 & 0.734{\tiny $\pm$ 0.000} & 0.646{\tiny $\pm$ 0.000} & \improve{-0.087} & 0.462{\tiny $\pm$ 0.000} & 0.558{\tiny $\pm$ 0.000} & \improve{+0.097} & 0.385{\tiny $\pm$ 0.000} & 0.331{\tiny $\pm$ 0.000} & \improve{-0.053} \\
 & SMART & 0.571 & 0.007 & 0.680{\tiny $\pm$ 0.001} & 0.647{\tiny $\pm$ 0.000} & \improve{-0.034} & 0.462{\tiny $\pm$ 0.000} & 0.558{\tiny $\pm$ 0.000} & \improve{+0.097} & 0.385{\tiny $\pm$ 0.000} & 0.332{\tiny $\pm$ 0.000} & \improve{-0.054} \\
\midrule
\multirow{6}{*}{\shortstack{MSLR-\\WEB10K}} & Uncal. & 0.394 & 0.292 & 1.387{\tiny $\pm$ 0.000} & 0.450{\tiny $\pm$ 0.001} & \improve{-0.937} & 0.653{\tiny $\pm$ 0.000} & 0.906{\tiny $\pm$ 0.000} & \improve{+0.252} & 0.605{\tiny $\pm$ 0.000} & 0.344{\tiny $\pm$ 0.000} & \improve{-0.261} \\
 & TS & 0.394 & 0.070 & 0.684{\tiny $\pm$ 0.000} & 0.448{\tiny $\pm$ 0.001} & \improve{-0.236} & 0.678{\tiny $\pm$ 0.000} & 0.906{\tiny $\pm$ 0.000} & \improve{+0.229} & 0.557{\tiny $\pm$ 0.000} & 0.343{\tiny $\pm$ 0.000} & \improve{-0.214} \\
 & DIAG & 0.394 & 0.096 & 0.651{\tiny $\pm$ 0.000} & 0.453{\tiny $\pm$ 0.001} & \improve{-0.198} & 0.743{\tiny $\pm$ 0.001} & 0.905{\tiny $\pm$ 0.000} & \improve{+0.162} & 0.466{\tiny $\pm$ 0.002} & 0.345{\tiny $\pm$ 0.000} & \improve{-0.120} \\
 & Spline & 0.394 & 0.011 & 0.663{\tiny $\pm$ 0.000} & 0.469{\tiny $\pm$ 0.000} & \improve{-0.193} & 0.673{\tiny $\pm$ 0.001} & 0.889{\tiny $\pm$ 0.001} & \improve{+0.216} & 0.547{\tiny $\pm$ 0.000} & 0.380{\tiny $\pm$ 0.001} & \improve{-0.167} \\
 & h-cal & 0.394 & 0.055 & 0.685{\tiny $\pm$ 0.000} & 0.448{\tiny $\pm$ 0.001} & \improve{-0.237} & 0.667{\tiny $\pm$ 0.000} & 0.906{\tiny $\pm$ 0.000} & \improve{+0.239} & 0.555{\tiny $\pm$ 0.000} & 0.343{\tiny $\pm$ 0.000} & \improve{-0.212} \\
 & SMART & 0.394 & 0.114 & 0.637{\tiny $\pm$ 0.001} & 0.456{\tiny $\pm$ 0.002} & \improve{-0.180} & 0.723{\tiny $\pm$ 0.004} & 0.902{\tiny $\pm$ 0.002} & \improve{+0.179} & 0.457{\tiny $\pm$ 0.002} & 0.357{\tiny $\pm$ 0.003} & \improve{-0.099} \\
\midrule
\multirow{6}{*}{\shortstack{Amazon\\ESCI}} & Uncal. & 0.764 & 0.047 & 0.528{\tiny $\pm$ 0.000} & 0.468{\tiny $\pm$ 0.001} & \improve{-0.060} & 0.391{\tiny $\pm$ 0.000} & 0.525{\tiny $\pm$ 0.002} & \improve{+0.133} & 0.161{\tiny $\pm$ 0.000} & 0.151{\tiny $\pm$ 0.000} & \improve{-0.010} \\
 & TS & 0.764 & 0.047 & 0.526{\tiny $\pm$ 0.000} & 0.468{\tiny $\pm$ 0.001} & \improve{-0.058} & 0.391{\tiny $\pm$ 0.000} & 0.524{\tiny $\pm$ 0.002} & \improve{+0.133} & 0.161{\tiny $\pm$ 0.000} & 0.151{\tiny $\pm$ 0.000} & \improve{-0.010} \\
 & DIAG & 0.764 & 0.045 & 0.525{\tiny $\pm$ 0.000} & 0.468{\tiny $\pm$ 0.001} & \improve{-0.057} & 0.391{\tiny $\pm$ 0.000} & 0.524{\tiny $\pm$ 0.002} & \improve{+0.133} & 0.161{\tiny $\pm$ 0.000} & 0.151{\tiny $\pm$ 0.000} & \improve{-0.010} \\
 & Spline & 0.764 & 0.019 & 0.516{\tiny $\pm$ 0.001} & 0.468{\tiny $\pm$ 0.001} & \improve{-0.048} & 0.391{\tiny $\pm$ 0.000} & 0.523{\tiny $\pm$ 0.003} & \improve{+0.132} & 0.161{\tiny $\pm$ 0.001} & 0.151{\tiny $\pm$ 0.001} & \improve{-0.010} \\
 & h-cal & 0.764 & 0.040 & 0.528{\tiny $\pm$ 0.003} & 0.468{\tiny $\pm$ 0.001} & \improve{-0.060} & 0.391{\tiny $\pm$ 0.000} & 0.524{\tiny $\pm$ 0.001} & \improve{+0.133} & 0.161{\tiny $\pm$ 0.000} & 0.151{\tiny $\pm$ 0.000} & \improve{-0.010} \\
 & SMART & 0.764 & 0.021 & 0.516{\tiny $\pm$ 0.000} & 0.468{\tiny $\pm$ 0.001} & \improve{-0.048} & 0.392{\tiny $\pm$ 0.000} & 0.525{\tiny $\pm$ 0.001} & \improve{+0.133} & 0.162{\tiny $\pm$ 0.001} & 0.152{\tiny $\pm$ 0.000} & \improve{-0.011} \\
\midrule
\multirow{6}{*}{SciDocs} & Uncal. & 0.562 & 0.210 & 0.913{\tiny $\pm$ 0.000} & 0.679{\tiny $\pm$ 0.001} & \improve{-0.234} & 0.480{\tiny $\pm$ 0.000} & 0.477{\tiny $\pm$ 0.004} & -0.002 & 0.398{\tiny $\pm$ 0.000} & 0.381{\tiny $\pm$ 0.001} & \improve{-0.017} \\
 & TS & 0.562 & 0.014 & 0.682{\tiny $\pm$ 0.000} & 0.679{\tiny $\pm$ 0.001} & \improve{-0.003} & 0.480{\tiny $\pm$ 0.000} & 0.476{\tiny $\pm$ 0.006} & -0.003 & 0.398{\tiny $\pm$ 0.000} & 0.382{\tiny $\pm$ 0.002} & \improve{-0.016} \\
 & DIAG & 0.562 & 0.008 & 0.682{\tiny $\pm$ 0.000} & 0.679{\tiny $\pm$ 0.001} & \improve{-0.003} & 0.480{\tiny $\pm$ 0.000} & 0.476{\tiny $\pm$ 0.007} & -0.004 & 0.398{\tiny $\pm$ 0.000} & 0.382{\tiny $\pm$ 0.003} & \improve{-0.016} \\
 & Spline & 0.562 & 0.015 & 0.682{\tiny $\pm$ 0.000} & 0.680{\tiny $\pm$ 0.000} & \improve{-0.002} & 0.478{\tiny $\pm$ 0.002} & 0.474{\tiny $\pm$ 0.004} & -0.005 & 0.397{\tiny $\pm$ 0.002} & 0.385{\tiny $\pm$ 0.001} & \improve{-0.012} \\
 & h-cal & 0.562 & 0.142 & 0.776{\tiny $\pm$ 0.000} & 0.679{\tiny $\pm$ 0.001} & \improve{-0.096} & 0.480{\tiny $\pm$ 0.000} & 0.477{\tiny $\pm$ 0.006} & -0.003 & 0.398{\tiny $\pm$ 0.000} & 0.383{\tiny $\pm$ 0.002} & \improve{-0.015} \\
 & SMART & 0.562 & 0.016 & 0.681{\tiny $\pm$ 0.000} & 0.680{\tiny $\pm$ 0.001} & \improve{-0.001} & 0.480{\tiny $\pm$ 0.000} & 0.477{\tiny $\pm$ 0.006} & -0.002 & 0.398{\tiny $\pm$ 0.000} & 0.381{\tiny $\pm$ 0.001} & \improve{-0.017} \\
\midrule
\multirow{6}{*}{WANDS} & Uncal. & 0.777 & 0.006 & 0.442{\tiny $\pm$ 0.000} & 0.436{\tiny $\pm$ 0.000} & \improve{-0.005} & 0.463{\tiny $\pm$ 0.000} & 0.490{\tiny $\pm$ 0.001} & \improve{+0.026} & 0.091{\tiny $\pm$ 0.000} & 0.088{\tiny $\pm$ 0.000} & \improve{-0.003} \\
 & TS & 0.777 & 0.006 & 0.442{\tiny $\pm$ 0.000} & 0.436{\tiny $\pm$ 0.000} & \improve{-0.005} & 0.463{\tiny $\pm$ 0.000} & 0.490{\tiny $\pm$ 0.001} & \improve{+0.026} & 0.091{\tiny $\pm$ 0.000} & 0.088{\tiny $\pm$ 0.000} & \improve{-0.003} \\
 & DIAG & 0.777 & 0.006 & 0.441{\tiny $\pm$ 0.000} & 0.436{\tiny $\pm$ 0.000} & \improve{-0.005} & 0.464{\tiny $\pm$ 0.000} & 0.490{\tiny $\pm$ 0.001} & \improve{+0.026} & 0.091{\tiny $\pm$ 0.000} & 0.088{\tiny $\pm$ 0.000} & \improve{-0.003} \\
 & Spline & 0.777 & 0.009 & 0.442{\tiny $\pm$ 0.000} & 0.437{\tiny $\pm$ 0.001} & \improve{-0.005} & 0.463{\tiny $\pm$ 0.001} & 0.489{\tiny $\pm$ 0.002} & \improve{+0.025} & 0.091{\tiny $\pm$ 0.000} & 0.088{\tiny $\pm$ 0.000} & \improve{-0.003} \\
 & h-cal & 0.777 & 0.009 & 0.442{\tiny $\pm$ 0.000} & 0.437{\tiny $\pm$ 0.001} & \improve{-0.005} & 0.464{\tiny $\pm$ 0.000} & 0.490{\tiny $\pm$ 0.001} & \improve{+0.026} & 0.091{\tiny $\pm$ 0.000} & 0.088{\tiny $\pm$ 0.000} & \improve{-0.003} \\
 & SMART & 0.777 & 0.006 & 0.441{\tiny $\pm$ 0.000} & 0.437{\tiny $\pm$ 0.001} & \improve{-0.005} & 0.463{\tiny $\pm$ 0.000} & 0.489{\tiny $\pm$ 0.001} & \improve{+0.026} & 0.091{\tiny $\pm$ 0.000} & 0.088{\tiny $\pm$ 0.000} & \improve{-0.003} \\
\midrule
\multirow{6}{*}{Alloprof-Rerank} & Uncal. & 0.757 & 0.150 & 0.662{\tiny $\pm$ 0.000} & 0.306{\tiny $\pm$ 0.001} & \improve{-0.357} & 0.443{\tiny $\pm$ 0.000} & 0.761{\tiny $\pm$ 0.000} & \improve{+0.318} & 0.099{\tiny $\pm$ 0.000} & 0.059{\tiny $\pm$ 0.000} & \improve{-0.040} \\
 & TS & 0.757 & 0.044 & 0.467{\tiny $\pm$ 0.000} & 0.300{\tiny $\pm$ 0.001} & \improve{-0.167} & 0.443{\tiny $\pm$ 0.000} & 0.761{\tiny $\pm$ 0.000} & \improve{+0.318} & 0.099{\tiny $\pm$ 0.000} & 0.059{\tiny $\pm$ 0.000} & \improve{-0.040} \\
 & DIAG & 0.757 & 0.044 & 0.467{\tiny $\pm$ 0.000} & 0.300{\tiny $\pm$ 0.001} & \improve{-0.167} & 0.443{\tiny $\pm$ 0.000} & 0.761{\tiny $\pm$ 0.000} & \improve{+0.318} & 0.099{\tiny $\pm$ 0.000} & 0.059{\tiny $\pm$ 0.000} & \improve{-0.040} \\
 & Spline & 0.757 & 0.038 & 0.470{\tiny $\pm$ 0.001} & 0.303{\tiny $\pm$ 0.001} & \improve{-0.168} & 0.424{\tiny $\pm$ 0.009} & 0.751{\tiny $\pm$ 0.003} & \improve{+0.327} & 0.101{\tiny $\pm$ 0.001} & 0.059{\tiny $\pm$ 0.000} & \improve{-0.041} \\
 & h-cal & 0.757 & 0.106 & 0.530{\tiny $\pm$ 0.000} & 0.302{\tiny $\pm$ 0.001} & \improve{-0.228} & 0.443{\tiny $\pm$ 0.000} & 0.761{\tiny $\pm$ 0.000} & \improve{+0.318} & 0.099{\tiny $\pm$ 0.000} & 0.059{\tiny $\pm$ 0.000} & \improve{-0.040} \\
 & SMART & 0.757 & 0.014 & 0.459{\tiny $\pm$ 0.000} & 0.300{\tiny $\pm$ 0.001} & \improve{-0.159} & 0.443{\tiny $\pm$ 0.000} & 0.761{\tiny $\pm$ 0.000} & \improve{+0.318} & 0.099{\tiny $\pm$ 0.000} & 0.059{\tiny $\pm$ 0.000} & \improve{-0.040} \\
\bottomrule
\end{tabular}
\end{table*}

%% file: table/budgeted_fallback.tex
\begin{table}[t]
\centering
\small
\caption{Budgeted fallback simulation. At automatic coverage \(\tau\), each reliability score keeps the lowest-risk \(\tau\) fraction for automatic use and routes the remaining highest-risk predictions to fallback or review. Values compare MRP with the no-projection confidence baseline \(\hat{q}_{\varnothing}^{\mathcal{A}}=\hat{c}^{\mathcal{A}}\). Positive values indicate a better reliability reranking.}
\label{tab:budgeted_fallback}
\setlength{\tabcolsep}{3.8pt}
\begin{tabular}{@{}p{0.31\linewidth}rrrr@{}}
\toprule
\multirow{2}{*}{Dataset} & \multicolumn{4}{c}{\(\Delta\mathrm{SelAcc}@\tau\)} \\
\cmidrule(l){2-5}
 & 10\% & 50\% & 70\% & 90\% \\
\midrule
ESCI-Rerank-US & \textbf{+0.043} & \textbf{+0.082} & \textbf{+0.056} & \textbf{+0.016} \\
MSLR-WEB10K & \textbf{+0.331} & \textbf{+0.174} & \textbf{+0.113} & \textbf{+0.028} \\
Amazon ESCI & \textbf{+0.003} & \textbf{+0.007} & \textbf{+0.014} & \textbf{+0.030} \\
SciDocs & \textbf{+0.074} & -0.003 & -0.003 & -0.001 \\
WANDS & \textbf{+0.005} & \textbf{+0.002} & \textbf{+0.004} & \textbf{+0.004} \\
Alloprof-Rerank & +0.000 & \textbf{+0.050} & \textbf{+0.101} & \textbf{+0.041} \\
\midrule
\textbf{Average} & \textbf{+0.076} & \textbf{+0.052} & \textbf{+0.047} & \textbf{+0.020} \\
\bottomrule
\end{tabular}
\end{table}

%% file: table/label_conditional_residual.tex
\begin{table}[t]
\centering
\small
\caption{Label-conditioned reliability spread. Within each calibrated-confidence group, we measure how much \(Z-\hat c^{\mathcal{A}}\) changes across predicted labels. A large spread indicates that equal-confidence predictions can have label-dependent correctness reliability. The random column is the 95th percentile after shuffling labels within each group.}
\label{tab:label_conditional_residual}
\setlength{\tabcolsep}{3.8pt}
\begin{tabular}{lrrrr}
\toprule
Dataset & Label spread & Random & Ratio & Seeds \\
\midrule
ESCI-Rerank-US & 26.7pp & 3.8pp & 6.9$\times$ & 18/18 \\
MSLR-WEB10K & 59.8pp & 4.3pp & 13.8$\times$ & 18/18 \\
Amazon ESCI & 60.8pp & 8.1pp & 7.5$\times$ & 18/18 \\
SciDocs & 6.6pp & 4.2pp & 1.6$\times$ & 18/18 \\
WANDS & 14.3pp & 3.0pp & 4.8$\times$ & 18/18 \\
Alloprof-Rerank & 57.3pp & 4.9pp & 11.6$\times$ & 18/18 \\
\bottomrule
\end{tabular}
\end{table}

%% file: table/mrp_ablation_results.tex
\begin{table}[t]
\centering
\small
\caption{Structural ablation averaged over all relevance-prediction settings. C-NLL denotes \(\mathrm{NLL}_{\mathrm{correct}}\). Variants are defined in Section~\ref{sec:ablation_variants}. MRP denotes the main label-wise 1D projection.}
\label{tab:mrp_ablation_results}
\setlength{\tabcolsep}{2.7pt}
\begin{tabular}{lrrrr}
\toprule
Variant & C-NLL$\downarrow$ & AUPR-Error$\uparrow$ & AURC$\downarrow$ & SelAcc@50$\uparrow$ \\
\midrule
Conf. & 0.6181 & 0.4876 & 0.2775 & 0.7279 \\
Shared 1D & 0.5720 & 0.4876 & 0.2775 & 0.7279 \\
Label-only intercept & 0.5132 & 0.6160 & 0.2302 & 0.7773 \\
Per-label isotonic & 0.5166 & 0.6148 & 0.2294 & 0.7782 \\
\textbf{MRP (Label-wise 1D)} & 0.4977 & \textbf{0.6185} & \textbf{0.2273} & \textbf{0.7800} \\
Label-wise 2D & \textbf{0.4975} & 0.6184 & 0.2274 & 0.7795 \\
\bottomrule
\end{tabular}
\end{table}

%% file: table/simplex_projection.tex
\begin{table*}[t]
\centering
\scriptsize
\setlength{\tabcolsep}{1.9pt}
\renewcommand{\arraystretch}{0.92}
\caption{Top-label simplex projection analysis. Base is the original calibrated vector, and MRC is the MRP-induced power-temperature projection. For Fixed ECE, Top ECE, NLL, and Brier, \(\Delta\) is MRC minus Base. Blue bold negative \(\Delta\) values indicate improvements, and lower is better for all four metrics. The \(q<1/K\) column reports infeasible targets. Solved gives exact rates.}
\label{tab:simplex_projection}
\newcommand{\mrcimprove}[1]{\textcolor{blue}{\textbf{#1}}}
\begin{tabular}{lclcccccccccccccc}
\toprule
\multirow{2}{*}{Dataset} & \multirow{2}{*}{\(K\)} & \multirow{2}{*}{Cal.} & \multicolumn{3}{c}{Fixed ECE\(\downarrow\)} & \multicolumn{3}{c}{Top ECE\(\downarrow\)} & \multicolumn{3}{c}{NLL\(\downarrow\)} & \multicolumn{3}{c}{Brier\(\downarrow\)} & \multirow{2}{*}{\(q<1/K\)} & \multirow{2}{*}{Solved} \\
\cmidrule(lr){4-6}\cmidrule(lr){7-9}\cmidrule(lr){10-12}\cmidrule(lr){13-15}
 & & & Base & MRC & \(\Delta\) & Base & MRC & \(\Delta\) & Base & MRC & \(\Delta\) & Base & MRC & \(\Delta\) & & \\
\midrule
\multirow{6}{*}{\shortstack{ESCI-Rerank\\US}} & \multirow{6}{*}{2} & Uncal. & 0.1835 & 0.0494 & \mrcimprove{-0.1341} & 0.1835 & 0.0494 & \mrcimprove{-0.1341} & 0.8323 & 0.6543 & \mrcimprove{-0.1780} & 0.5722 & 0.4620 & \mrcimprove{-0.1103} & 37.6\% & 100.0\% \\
 &  & TS & 0.0131 & 0.0482 & +0.0351 & 0.0131 & 0.0482 & +0.0351 & 0.6793 & 0.6541 & \mrcimprove{-0.0252} & 0.4863 & 0.4618 & \mrcimprove{-0.0245} & 37.6\% & 100.0\% \\
 &  & DIAG & 0.0111 & 0.0476 & +0.0365 & 0.0111 & 0.0476 & +0.0365 & 0.6795 & 0.6538 & \mrcimprove{-0.0257} & 0.4865 & 0.4616 & \mrcimprove{-0.0249} & 37.7\% & 100.0\% \\
 &  & Spline & 0.0153 & 0.0413 & +0.0260 & 0.0114 & 0.0452 & +0.0338 & 0.6802 & 0.6558 & \mrcimprove{-0.0244} & 0.4872 & 0.4635 & \mrcimprove{-0.0237} & 37.7\% & 91.7\% \\
 &  & h-cal & 0.1122 & 0.0506 & \mrcimprove{-0.0616} & 0.1122 & 0.0506 & \mrcimprove{-0.0616} & 0.7338 & 0.6541 & \mrcimprove{-0.0797} & 0.5242 & 0.4618 & \mrcimprove{-0.0624} & 37.4\% & 100.0\% \\
 &  & SMART & 0.0074 & 0.0478 & +0.0404 & 0.0074 & 0.0478 & +0.0404 & 0.6802 & 0.6543 & \mrcimprove{-0.0259} & 0.4872 & 0.4620 & \mrcimprove{-0.0252} & 38.5\% & 100.0\% \\
\midrule
\multirow{6}{*}{\shortstack{MSLR-\\WEB10K}} & \multirow{6}{*}{5} & Uncal. & 0.2922 & 0.0703 & \mrcimprove{-0.2219} & 0.2922 & 0.0762 & \mrcimprove{-0.2160} & 2.1775 & 1.2081 & \mrcimprove{-0.9694} & 0.8695 & 0.6274 & \mrcimprove{-0.2422} & 31.9\% & 100.0\% \\
 &  & TS & 0.0697 & 0.0681 & \mrcimprove{-0.0016} & 0.0697 & 0.0750 & +0.0052 & 1.4052 & 1.2090 & \mrcimprove{-0.1963} & 0.7302 & 0.6262 & \mrcimprove{-0.1039} & 32.1\% & 100.0\% \\
 &  & DIAG & 0.0964 & 0.0785 & \mrcimprove{-0.0178} & 0.0964 & 0.0857 & \mrcimprove{-0.0107} & 1.3380 & 1.3085 & \mrcimprove{-0.0294} & 0.7156 & 0.6321 & \mrcimprove{-0.0835} & 32.0\% & 100.0\% \\
 &  & Spline & 0.0107 & 0.0545 & +0.0438 & 0.0950 & 0.1151 & +0.0201 & 1.4763 & 1.4506 & \mrcimprove{-0.0256} & 0.7481 & 0.6973 & \mrcimprove{-0.0508} & 31.9\% & 57.9\% \\
 &  & h-cal & 0.0549 & 0.0679 & +0.0130 & 0.0548 & 0.0747 & +0.0199 & 1.4231 & 1.2307 & \mrcimprove{-0.1924} & 0.7330 & 0.6316 & \mrcimprove{-0.1014} & 32.0\% & 99.9\% \\
 &  & SMART & 0.1137 & 0.0709 & \mrcimprove{-0.0428} & 0.1137 & 0.0772 & \mrcimprove{-0.0365} & 1.3301 & 1.2178 & \mrcimprove{-0.1123} & 0.6998 & 0.6296 & \mrcimprove{-0.0702} & 32.0\% & 100.0\% \\
\midrule
\multirow{6}{*}{\shortstack{Amazon\\ESCI}} & \multirow{6}{*}{4} & Uncal. & 0.0470 & 0.0136 & \mrcimprove{-0.0334} & 0.0470 & 0.0194 & \mrcimprove{-0.0276} & 0.7132 & 0.7191 & +0.0060 & 0.3708 & 0.3496 & \mrcimprove{-0.0212} & 7.2\% & 100.0\% \\
 &  & TS & 0.0469 & 0.0135 & \mrcimprove{-0.0334} & 0.0469 & 0.0194 & \mrcimprove{-0.0275} & 0.7107 & 0.7192 & +0.0085 & 0.3710 & 0.3496 & \mrcimprove{-0.0213} & 7.2\% & 100.0\% \\
 &  & DIAG & 0.0453 & 0.0134 & \mrcimprove{-0.0318} & 0.0453 & 0.0193 & \mrcimprove{-0.0260} & 0.7103 & 0.7230 & +0.0126 & 0.3709 & 0.3497 & \mrcimprove{-0.0213} & 7.2\% & 100.0\% \\
 &  & Spline & 0.0188 & 0.0122 & \mrcimprove{-0.0066} & 0.0178 & 0.0177 & 0.0000 & 0.7012 & 0.7078 & +0.0066 & 0.3672 & 0.3497 & \mrcimprove{-0.0176} & 7.0\% & 98.8\% \\
 &  & h-cal & 0.0400 & 0.0137 & \mrcimprove{-0.0263} & 0.0400 & 0.0192 & \mrcimprove{-0.0208} & 0.7134 & 0.7090 & \mrcimprove{-0.0043} & 0.3700 & 0.3495 & \mrcimprove{-0.0205} & 7.1\% & 100.0\% \\
 &  & SMART & 0.0206 & 0.0140 & \mrcimprove{-0.0066} & 0.0206 & 0.0197 & \mrcimprove{-0.0010} & 0.7024 & 0.7192 & +0.0167 & 0.3674 & 0.3496 & \mrcimprove{-0.0179} & 7.2\% & 100.0\% \\
\midrule
\multirow{6}{*}{SciDocs} & \multirow{6}{*}{2} & Uncal. & 0.2097 & 0.0088 & \mrcimprove{-0.2008} & 0.2097 & 0.0103 & \mrcimprove{-0.1993} & 0.9131 & 0.6788 & \mrcimprove{-0.2343} & 0.5962 & 0.4860 & \mrcimprove{-0.1102} & 13.0\% & 100.0\% \\
 &  & TS & 0.0139 & 0.0074 & \mrcimprove{-0.0065} & 0.0139 & 0.0078 & \mrcimprove{-0.0061} & 0.6818 & 0.6789 & \mrcimprove{-0.0029} & 0.4887 & 0.4862 & \mrcimprove{-0.0025} & 8.6\% & 100.0\% \\
 &  & DIAG & 0.0081 & 0.0070 & \mrcimprove{-0.0010} & 0.0081 & 0.0073 & \mrcimprove{-0.0008} & 0.6815 & 0.6788 & \mrcimprove{-0.0027} & 0.4884 & 0.4861 & \mrcimprove{-0.0023} & 9.0\% & 100.0\% \\
 &  & Spline & 0.0152 & 0.0127 & \mrcimprove{-0.0024} & 0.0136 & 0.0103 & \mrcimprove{-0.0033} & 0.6819 & 0.6802 & \mrcimprove{-0.0017} & 0.4889 & 0.4872 & \mrcimprove{-0.0016} & 6.8\% & 89.0\% \\
 &  & h-cal & 0.1417 & 0.0054 & \mrcimprove{-0.1363} & 0.1417 & 0.0059 & \mrcimprove{-0.1358} & 0.7755 & 0.6791 & \mrcimprove{-0.0964} & 0.5434 & 0.4864 & \mrcimprove{-0.0571} & 10.9\% & 100.0\% \\
 &  & SMART & 0.0162 & 0.0119 & \mrcimprove{-0.0043} & 0.0162 & 0.0126 & \mrcimprove{-0.0036} & 0.6812 & 0.6797 & \mrcimprove{-0.0016} & 0.4882 & 0.4868 & \mrcimprove{-0.0014} & 8.8\% & 100.0\% \\
\midrule
\multirow{6}{*}{WANDS} & \multirow{6}{*}{3} & Uncal. & 0.0059 & 0.0077 & +0.0018 & 0.0059 & 0.0081 & +0.0022 & 0.5506 & 0.5517 & +0.0011 & 0.3140 & 0.3124 & \mrcimprove{-0.0016} & 1.0\% & 100.0\% \\
 &  & TS & 0.0058 & 0.0077 & +0.0019 & 0.0058 & 0.0080 & +0.0022 & 0.5506 & 0.5517 & +0.0011 & 0.3140 & 0.3124 & \mrcimprove{-0.0016} & 1.0\% & 100.0\% \\
 &  & DIAG & 0.0058 & 0.0075 & +0.0017 & 0.0058 & 0.0078 & +0.0020 & 0.5504 & 0.5521 & +0.0017 & 0.3140 & 0.3124 & \mrcimprove{-0.0016} & 1.0\% & 100.0\% \\
 &  & Spline & 0.0086 & 0.0089 & +0.0004 & 0.0078 & 0.0085 & +0.0007 & 0.5509 & 0.5518 & +0.0009 & 0.3143 & 0.3129 & \mrcimprove{-0.0015} & 1.2\% & 98.1\% \\
 &  & h-cal & 0.0095 & 0.0082 & \mrcimprove{-0.0013} & 0.0095 & 0.0086 & \mrcimprove{-0.0009} & 0.5506 & 0.5522 & +0.0017 & 0.3140 & 0.3125 & \mrcimprove{-0.0016} & 1.1\% & 100.0\% \\
 &  & SMART & 0.0061 & 0.0080 & +0.0018 & 0.0061 & 0.0083 & +0.0022 & 0.5503 & 0.5518 & +0.0015 & 0.3140 & 0.3125 & \mrcimprove{-0.0015} & 1.0\% & 100.0\% \\
\midrule
\multirow{6}{*}{\shortstack{Alloprof-\\Rerank}} & \multirow{6}{*}{2} & Uncal. & 0.1501 & 0.0730 & \mrcimprove{-0.0771} & 0.1501 & 0.0757 & \mrcimprove{-0.0744} & 0.6623 & 0.3447 & \mrcimprove{-0.3176} & 0.3752 & 0.2155 & \mrcimprove{-0.1597} & 26.3\% & 100.0\% \\
 &  & TS & 0.0437 & 0.0740 & +0.0303 & 0.0437 & 0.0725 & +0.0288 & 0.4668 & 0.3398 & \mrcimprove{-0.1270} & 0.3134 & 0.2141 & \mrcimprove{-0.0993} & 24.9\% & 100.0\% \\
 &  & DIAG & 0.0439 & 0.0746 & +0.0307 & 0.0439 & 0.0732 & +0.0293 & 0.4670 & 0.3398 & \mrcimprove{-0.1272} & 0.3136 & 0.2141 & \mrcimprove{-0.0995} & 24.9\% & 100.0\% \\
 &  & Spline & 0.0380 & 0.0704 & +0.0324 & 0.0370 & 0.0676 & +0.0306 & 0.4704 & 0.3452 & \mrcimprove{-0.1251} & 0.3130 & 0.2181 & \mrcimprove{-0.0949} & 23.9\% & 97.1\% \\
 &  & h-cal & 0.1055 & 0.0755 & \mrcimprove{-0.0301} & 0.1055 & 0.0741 & \mrcimprove{-0.0315} & 0.5298 & 0.3417 & \mrcimprove{-0.1881} & 0.3418 & 0.2147 & \mrcimprove{-0.1270} & 24.8\% & 100.0\% \\
 &  & SMART & 0.0143 & 0.0745 & +0.0602 & 0.0143 & 0.0740 & +0.0598 & 0.4589 & 0.3400 & \mrcimprove{-0.1189} & 0.3079 & 0.2140 & \mrcimprove{-0.0939} & 25.2\% & 100.0\% \\
\bottomrule
\end{tabular}
\renewcommand{\arraystretch}{1.0}
\end{table*}

%% file: references.bib
@article{liu2009learning,
  author  = {Liu, Tie-Yan},
  title   = {Learning to Rank for Information Retrieval},
  journal = {Foundations and Trends in Information Retrieval},
  volume  = {3},
  number  = {3},
  pages   = {225--331},
  year    = {2009},
  doi     = {10.1561/1500000016}
}

@article{reddy2022shopping,
  title={Shopping queries dataset: A large-scale ESCI benchmark for improving product search},
  author={Reddy, Chandan K and M{\`a}rquez, Llu{\'\i}s and Valero, Fran and Rao, Nikhil and Zaragoza, Hugo and Bandyopadhyay, Sambaran and Biswas, Arnab and Xing, Anlu and Subbian, Karthik},
  journal={arXiv preprint arXiv:2206.06588},
  year={2022}
}

@inproceedings{chen2022wands,
  title={Wands: Dataset for product search relevance assessment},
  author={Chen, Yan and Liu, Shujian and Liu, Zheng and Sun, Weiyi and Baltrunas, Linas and Schroeder, Benjamin},
  booktitle={European Conference on Information Retrieval},
  pages={128--141},
  year={2022},
  organization={Springer}
}

@article{lewis2020retrieval,
  title={Retrieval-augmented generation for knowledge-intensive nlp tasks},
  author={Lewis, Patrick and Perez, Ethan and Piktus, Aleksandra and Petroni, Fabio and Karpukhin, Vladimir and Goyal, Naman and K{\"u}ttler, Heinrich and Lewis, Mike and Yih, Wen-tau and Rockt{\"a}schel, Tim and others},
  journal={Advances in neural information processing systems},
  volume={33},
  pages={9459--9474},
  year={2020}
}

@article{nogueira2019passage,
  title={Passage Re-ranking with BERT},
  author={Nogueira, Rodrigo and Cho, Kyunghyun},
  journal={arXiv preprint arXiv:1901.04085},
  year={2019}
}

@inproceedings{karpukhin2020dense,
  title={Dense passage retrieval for open-domain question answering},
  author={Karpukhin, Vladimir and Oguz, Barlas and Min, Sewon and Lewis, Patrick and Wu, Ledell and Edunov, Sergey and Chen, Danqi and Yih, Wen-tau},
  booktitle={Proceedings of the 2020 conference on empirical methods in natural language processing (EMNLP)},
  pages={6769--6781},
  year={2020}
}

@inproceedings{guu2020retrieval,
  title={Retrieval augmented language model pre-training},
  author={Guu, Kelvin and Lee, Kenton and Tung, Zora and Pasupat, Panupong and Chang, Mingwei},
  booktitle={International conference on machine learning},
  pages={3929--3938},
  year={2020},
  organization={PMLR}
}

@inproceedings{izacard2021leveraging,
  title={Leveraging passage retrieval with generative models for open domain question answering},
  author={Izacard, Gautier and Grave, Edouard},
  booktitle={Proceedings of the 16th conference of the european chapter of the association for computational linguistics: main volume},
  pages={874--880},
  year={2021}
}

@article{geifman2017selective,
  title={Selective classification for deep neural networks},
  author={Geifman, Yonatan and El-Yaniv, Ran},
  journal={Advances in neural information processing systems},
  volume={30},
  year={2017}
}

@inproceedings{zadrozny2002transforming,
  title={Transforming classifier scores into accurate multiclass probability estimates},
  author={Zadrozny, Bianca and Elkan, Charles},
  booktitle={Proceedings of the eighth ACM SIGKDD international conference on Knowledge discovery and data mining},
  pages={694--699},
  year={2002}
}

@inproceedings{guo2017calibration,
  title={On calibration of modern neural networks},
  author={Guo, Chuan and Pleiss, Geoff and Sun, Yu and Weinberger, Kilian Q},
  booktitle={International conference on machine learning},
  pages={1321--1330},
  year={2017},
  organization={PMLR}
}

@inproceedings{gupta2021top,
  title={Top-label calibration and multiclass-to-binary reductions},
  author={Gupta, Chirag and Ramdas, Aaditya},
  booktitle = {International Conference on Learning Representations},
  year      = {2022}
}

@article{platt1999probabilistic,
  title={Probabilistic outputs for support vector machines and comparisons to regularized likelihood methods},
  author={Platt, John and others},
  journal={Advances in large margin classifiers},
  volume={10},
  number={3},
  pages={61--74},
  year={1999},
  publisher={Cambridge, MA}
}

@inproceedings{niculescu2005predicting,
  title={Predicting good probabilities with supervised learning},
  author={Niculescu-Mizil, Alexandru and Caruana, Rich},
  booktitle={Proceedings of the 22nd international conference on Machine learning},
  pages={625--632},
  year={2005}
}

@article{kull2019beyond,
  title={Beyond temperature scaling: Obtaining well-calibrated multi-class probabilities with dirichlet calibration},
  author={Kull, Meelis and Perello Nieto, Miquel and K{\"a}ngsepp, Markus and Silva Filho, Telmo and Song, Hao and Flach, Peter},
  journal={Advances in neural information processing systems},
  volume={32},
  year={2019}
}

@inproceedings{kumar2018trainable,
  title={Trainable calibration measures for neural networks from kernel mean embeddings},
  author={Kumar, Aviral and Sarawagi, Sunita and Jain, Ujjwal},
  booktitle={International Conference on Machine Learning},
  pages={2805--2814},
  year={2018},
  organization={PMLR}
}

@article{el2010foundations,
  title={On the Foundations of Noise-free Selective Classification.},
  author={El-Yaniv, Ran and others},
  journal={Journal of Machine Learning Research},
  volume={11},
  number={5},
  year={2010}
}

@inproceedings{geifman2019selectivenet,
  title={Selectivenet: A deep neural network with an integrated reject option},
  author={Geifman, Yonatan and El-Yaniv, Ran},
  booktitle={International conference on machine learning},
  pages={2151--2159},
  year={2019},
  organization={PMLR}
}

@article{hendrycks2016baseline,
  title={A baseline for detecting misclassified and out-of-distribution examples in neural networks},
  author={Hendrycks, Dan and Gimpel, Kevin},
  journal={arXiv preprint arXiv:1610.02136},
  year={2016}
}

@article{corbiere2019addressing,
  title={Addressing failure prediction by learning model confidence},
  author={Corbi{\`e}re, Charles and Thome, Nicolas and Bar-Hen, Avner and Cord, Matthieu and P{\'e}rez, Patrick},
  journal={Advances in neural information processing systems},
  volume={32},
  year={2019}
}

@misc{microsoft2010mslr,
  author       = {{Microsoft Research}},
  title        = {{Microsoft Learning to Rank Datasets}},
  howpublished = {\url{https://www.microsoft.com/en-us/research/project/mslr/}},
  year         = {2010},
  note         = {Accessed: 2026-05-19}
}

@article{lefebvre2023alloprof,
  title={Alloprof: a new french question-answer education dataset and its use in an information retrieval case study},
  author={Lefebvre-Brossard, Antoine and Gazaille, Stephane and Desmarais, Michel C},
  journal={arXiv preprint arXiv:2302.07738},
  year={2023}
}

@inproceedings{cohan2020specter,
  title={Specter: Document-level representation learning using citation-informed transformers},
  author={Cohan, Arman and Feldman, Sergey and Beltagy, Iz and Downey, Doug and Weld, Daniel S},
  booktitle={Proceedings of the 58th annual meeting of the association for computational linguistics},
  pages={2270--2282},
  year={2020}
}

@article{thakur2021beir,
  title={Beir: A heterogenous benchmark for zero-shot evaluation of information retrieval models},
  author={Thakur, Nandan and Reimers, Nils and R{\"u}ckl{\'e}, Andreas and Srivastava, Abhishek and Gurevych, Iryna},
  journal={arXiv preprint arXiv:2104.08663},
  year={2021}
}

@article{gupta2020calibration,
  title={Calibration of neural networks using splines},
  author={Gupta, Kartik and Rahimi, Amir and Ajanthan, Thalaiyasingam and Mensink, Thomas and Sminchisescu, Cristian and Hartley, Richard},
  journal={arXiv preprint arXiv:2006.12800},
  year={2020}
}

@article{rahimi2020intra,
  title={Intra order-preserving functions for calibration of multi-class neural networks},
  author={Rahimi, Amir and Shaban, Amirreza and Cheng, Ching-An and Hartley, Richard and Boots, Byron},
  journal={Advances in neural information processing systems},
  volume={33},
  pages={13456--13467},
  year={2020}
}

@article{huang2025h,
  title={h-calibration: Rethinking Classifier Recalibration with Probabilistic Error-Bounded Objective},
  author={Huang, Wenjian and Cao, Guiping and Xia, Jiahao and Chen, Jingkun and Wang, Hao and Zhang, Jianguo},
  journal={IEEE Transactions on Pattern Analysis and Machine Intelligence},
  year={2025},
  publisher={IEEE}
}

@article{guo2025sample,
  title={Sample Margin-Aware Recalibration of Temperature Scaling},
  author={Guo, Haolan and Tao, Linwei and Luo, Haoyang and Dong, Minjing and Xu, Chang},
  journal={arXiv preprint arXiv:2506.23492},
  year={2025}
}

@inproceedings{naeini2015obtaining,
  title={Obtaining well calibrated probabilities using bayesian binning},
  author={Naeini, Mahdi Pakdaman and Cooper, Gregory and Hauskrecht, Milos},
  booktitle={Proceedings of the AAAI conference on artificial intelligence},
  volume={29},
  number={1},
  year={2015}
}

@article{kumar2019verified,
  title={Verified uncertainty calibration},
  author={Kumar, Ananya and Liang, Percy S and Ma, Tengyu},
  journal={Advances in neural information processing systems},
  volume={32},
  year={2019}
}

@inproceedings{vaicenavicius2019evaluating,
  title={Evaluating model calibration in classification},
  author={Vaicenavicius, Juozas and Widmann, David and Andersson, Carl and Lindsten, Fredrik and Roll, Jacob and Sch{\"o}n, Thomas},
  booktitle={The 22nd international conference on artificial intelligence and statistics},
  pages={3459--3467},
  year={2019},
  organization={PMLR}
}

@inproceedings{cortes2016learning,
  title={Learning with rejection},
  author={Cortes, Corinna and DeSalvo, Giulia and Mohri, Mehryar},
  booktitle={International conference on algorithmic learning theory},
  pages={67--82},
  year={2016},
  organization={Springer}
}

@article{madras2018predict,
  title={Predict responsibly: improving fairness and accuracy by learning to defer},
  author={Madras, David and Pitassi, Toni and Zemel, Richard},
  journal={Advances in neural information processing systems},
  volume={31},
  year={2018}
}

@inproceedings{mozannar2020consistent,
  title={Consistent estimators for learning to defer to an expert},
  author={Mozannar, Hussein and Sontag, David},
  booktitle={International conference on machine learning},
  pages={7076--7087},
  year={2020},
  organization={PMLR}
}

@inproceedings{craswell2021ms,
  title={Ms marco: Benchmarking ranking models in the large-data regime},
  author={Craswell, Nick and Mitra, Bhaskar and Yilmaz, Emine and Campos, Daniel and Lin, Jimmy},
  booktitle={Proceedings of the 44th international ACM SIGIR conference on research and development in information retrieval},
  pages={1566--1576},
  year={2021}
}

@book{robertson2009probabilistic,
  title={The probabilistic relevance framework: BM25 and beyond},
  author={Robertson, Stephen and Zaragoza, Hugo},
  volume={4},
  year={2009},
  publisher={Now Publishers Inc}
}

@article{heiss2026jucal,
  title={JUCAL: Jointly Calibrating Aleatoric and Epistemic Uncertainty in Classification Tasks},
  author={Heiss, Jakob and Lambrecht, S{\"o}ren and Weissteiner, Jakob and Wutte, Hanna and {\v{Z}}uri{\v{c}}, {\v{Z}}an and Teichmann, Josef and Yu, Bin},
  journal={arXiv preprint arXiv:2602.20153},
  year={2026}
}
